\documentclass[aps,prl,preprint,superscriptaddress,tightenlines]{revtex4}

\usepackage{graphicx}
\usepackage{dcolumn}
\usepackage{bm}

\begin{document}

\preprint{\vbox{\hbox{\hfil CLNS 02/1785}
                        \hbox{\hfil CLEO 02-06}
}}

\newcommand{\beq}{\begin{equation}}
\newcommand{\eeq}{\end{equation}}
\newcommand{\beqa}{\begin{eqnarray}}
\newcommand{\eeqa}{\end{eqnarray}}
\newcommand{\bit}{\begin{itemize}}
\newcommand{\eit}{\end{itemize}}
\newcommand{\p}{\psi}
\newcommand{\po}{\psi(1S)}
\newcommand{\pt}{\psi(2S)}
\newcommand{\co}{\chi_{c1}}
\newcommand{\ct}{\chi_{c2}}
\newcommand{\cost}{\cos\theta}
\newcommand{\vobs}{\vec{v}^{obs}}
\newcommand{\vgen}{\vec{v}^{gen}}
\newcommand{\Mmat}{{\bf M}}
\newcommand{\vv}{\vec{v}}
\newcommand{\MSE}{{\bf M}^{SE}}
\newcommand{\ME}{{\bf M}^{E}}
\newcommand{\MS}{{\bf M}^{S}}
\newcommand{\apo}{\alpha_{\psi(1S)}}
\newcommand{\apt}{\alpha_{\psi(2S)}}
\newcommand{\dels}{\not \! \partial}
\newcommand{\dslash}{\!\! \not \!\! D}
\newcommand{\OTS}{\langle{\cal O}_1(^3S_1)\rangle}
\newcommand{\ETS}{\langle{\cal O}_8(^3S_1)\rangle}
\newcommand{\ETP}{\langle{\cal O}_8(^3P_J)\rangle}
\newcommand{\ETPO}{\langle{\cal O}_8(^3P_0)\rangle}
\newcommand{\EOP}{\langle{\cal O}_8(^1P_1)\rangle}
\newcommand{\EOS}{\langle{\cal O}_8(^1S_0)\rangle}
\newcommand{\KOTS}{K_1(^3S_1)}
\newcommand{\KETS}{K_8(^3S_1)}
\newcommand{\KETP}{K_8(^3P_J)}
\newcommand{\KEOP}{K_8(^1P_1)}
\newcommand{\KEOS}{K_8(^1S_0)}
\newcommand{\s}{\scriptstyle}
\newcommand{\B}{$\bullet$}
\newcommand{\U}{$\circ$}

\title{Measurements of Inclusive $B \to \psi$ Production}

\author{S.~Anderson}
\author{V.~V.~Frolov}
\author{Y.~Kubota}
\author{S.~J.~Lee}
\author{S.~Z.~Li}
\author{R.~Poling}
\author{A.~Smith}
\author{C.~J.~Stepaniak}
\author{J.~Urheim}
\affiliation{University of Minnesota, Minneapolis, Minnesota 55455} 
\author{Z.~Metreveli}
\author{K.K.~Seth}
\author{A.~Tomaradze}
\author{P.~Zweber}
\affiliation{Northwestern University, Evanston, Illinois 60208}
% ...
\collaboration{CLEO Collaboration}
\noaffiliation
 
\date{May 5, 2002}

\begin{abstract}
Using the combined CLEO II and CLEO II.V data sets of 9.1 fb$^{-1}$ at 
the $\Upsilon(4S)$, we measure properties of $\psi$ mesons produced
directly from decays of the $B$ meson, 
where ``$B$'' denotes an admixture of 
$B^+$, $B^-$, $B^0$, and $\bar{B^0}$, 
and ``$\psi$'' denotes either $J/\po$ or $\pt$.
We report first measurements of $\psi$ polarization in
$B \to \psi \mbox{(direct)} X$:
$\alpha_{\po} = -0.30^{+0.07}_{-0.06}\pm 0.04$ and
$\alpha_{\pt} = -0.45^{+0.22}_{-0.19} \pm 0.04$.
We also report improved measurements of the momentum distributions of
$\psi$ produced directly from $B$ decays, correcting for
measurement smearing. 
Finally, we report measurements of the inclusive branching fraction for 
$B \to \psi X$ and $B \to \co X$.  
\end{abstract}

\maketitle

 
Inclusive production of $\psi$ is currently understood in the framework
of Non-Relativistic QCD (NRQCD) effective field theory \cite{NRQCD}.
In 1995, measurements of prompt $\psi$ production at the Tevatron 
\cite{CDFproduction,CDFproduction2} 
ruled out the then-dominant Color Singlet Model (CSM);  
in contrast, NRQCD calculations \cite{NRQCD-CDFproduction} 
could accommodate the relatively large production rate.
However, measurements of the polarization of these prompt $\psi$ 
\cite{CDFpolarization} deviated from NRQCD calculations at high $p_T$. 
The precision of these calculations is limited by the knowledge of the
process-independent, long-distance matrix elements (LDME's), which also 
appear in NRQCD calculations of $\psi$ production in $B$ decays.
The polarization of $\psi$ produced from $B$ decays 
\cite{Ma00alpha,Fleming97} is sensitive to the 
color-octet LDME's;  
however, these calculations have been done only to leading order (LO).
The momentum distribution of $\psi$ produced in $B$ decays 
\cite{Beneke00,Beneke97,Palmer97} is also 
sensitive to the dominant color octet terms, particularly at low $p_\psi$.  
Additionally, the low-momentum region would also be
affected by the existence of an intrinsic charm component
in $B$ mesons \cite{Chang01}.
Finally, the inclusive branching fraction ${\cal B}(B \to \psi X)$ 
\cite{Beneke97,Ma00,Ko96} constrains a sum of LDME's.
This Letter reports measurements of these three properties of 
$\psi$ production in $B$ decays, which could significantly 
improve the knowledge of the non-perturbative parameters of NRQCD. 

Our analysis \cite{MyThesis} is based on 9.7 million $B\overline{B}$ events
($9.1 \, \hbox{fb}^{-1}$) produced on the $\Upsilon(4S)$ resonance.
Additionally, $4.4 \, \hbox{fb}^{-1}$ of data collected slightly
below the $\Upsilon(4S)$ resonance were used 
to subtract the small $(\approx 2\%)$ contribution of continuum 
($e^+ e^- \to q\overline{q}, q \in \{ u,d,c,s\} $)
$\po$ production.
The $e^+ e^-$ collisions were delivered by the Cornell Electron Stoarge
Ring (CESR) and detected with two configurations of the CLEO detector, 
CLEO II \cite{CLEOII} and CLEO II.V \cite{CLEOIIV}.

We select events that have spherical energy distributions and are likely
to be hadronic.  We reconstruct $\psi$ candidates in the dilepton modes
$\psi \to \mu^+ \mu^-$ and $\psi \to e^+(\gamma) e^-(\gamma)$.
The selection criteria were chosen with a goal of high detection efficiency.
In the di-muon channel, 
at least one of the muon candidates must penetrate 
at least 3 interaction lengths into the iron of the solenoid return yoke; 
if only one candidate satisfies this, 
then the other candidate must leave a shower in the crystal calorimeter 
which is consistent with that of a minimum ionizing particle.
In the di-electron channel, 
we use shower information from the crystal calorimeter
and measurements of specific ionization from the drift
chamber to identify electron candidates.
We also attempt to recover up to one Bremsstrahlung photon 
for each electron candidate.  
To do this, we select the most collinear shower
within a five-degree cone around the initial electron direction;  
furthermore, the shower must not be associated with any track, and,
when combined with any other shower in the event, 
must not result in an invariant mass consistent with a $\pi^0$.
The PDG 2001 \cite{PDG2001} branching fractions are used to combine
results from the electron and muon channels,
except for ${\cal B}(\pt \to \mu^+ \mu^-)$, which we assume by 
lepton universality to be equal to ${\cal B}(\pt \to e^+ e^-)$,
with an uncertainty of 20\% of itself;  this is
consistent with recent measurements \cite{BaBarPT}.

About 30\% of $\po$ from $B$ decays 
have intermediate parents of $\pt$ or $\co$ \cite{PDG2001}.  
Our measurements of directly produced $\po$ are obtained by subtracting the 
contributions of these ``feed-down'' $\po$ from the inclusive $\po$ sample. 
For every event with a $\po$ candidate within $^{+25}_{-50}$ MeV of the
nominal mass, we search for these intermediate parents through 
the decay chains $\co \to \po \gamma$ and
$\pt \to \po \pi^+ \pi^-$.  
We reconstruct $(M_{\ell^+ \ell^- \pi^+ \pi^-}-M_{\ell^+ \ell^-})$ 
and $(M_{\ell^+ \ell^- \gamma}-M_{\ell^+ \ell^-})$,
which have better resolution than the reconstructed
$\pt$ and $\co$ invariant masses themselves.
In the $\pt \to \po \pi^+ \pi^-$ decay chain, we reduce low-momentum
pion background by requiring $M_{\pi^+\pi^-} > 0.45$ GeV, 
which has a efficiency of about 85\%, from Monte Carlo simulation.
This decay mode is also used to improve the statistics 
in our measurements of the inclusive branching fraction 
and the $\pt$ momentum distribution in $B \to \pt X$.
We do not reconstruct the related decay $\pt \to \po \pi^0 \pi^0$, and
argue that the properties of $\po$ from this decay are identical to those
of $\po$ from $\pt \to \po \pi^+ \pi^-$;
the kinematic difference in the momentum distribution is small
compared to the experimental resolution, and the isospin state of the
$\pi\pi$ state has no bearing on the polarization of the $\po$.

The CLEO Monte Carlo simulation, based on GEANT \cite{GEANT}, 
is used to obtain the invariant mass lineshape for signal events
and to estimate the detection efficiency.  
In these simulated signal events, one of the $B$ mesons decays via 
one of the decay chains listed above.
For each decay chain, we generate two samples of events;  
one with all $\psi$ longitudinally polarized, 
the other with all $\psi$ transversely polarized.  
We find that the detection efficiency varies slightly as a function of 
$\psi$ momentum and polarization.


The procedure and results for the inclusive branching fraction and 
momentum distribution measurements are as follows.
We divide the data into partitions in $p_\psi$, the
momentum of the $\psi$ candidate, using a binsize of 0.1 GeV/$c$.  
For each partition, the invariant mass distribution of $\psi$ candidates
is fit to a sum of 
a signal lineshape, obtained from the Monte Carlo simulation, and
a cubic polynomial background.   
The average $\chi^2$ of the fits 
is consistent with the number of degrees of freedom, 
thus justifying our choice of the above parametrization.
We repeat this procedure using signal Monte Carlo events,
binning in generated $\psi$ momentum, 
to obtain detection efficiencies as a function of $p_\psi$.
The data is then corrected for detection efficiency bin by bin;
this minimizes the effect of any discrepancy between the
true $p_\psi$ distribution and that generated by the Monte Carlo simulation.
Similarly, we fit the invariant mass distributions of
$(M_{\ell^+ \ell^- \pi^+ \pi^-}-M_{\ell^+ \ell^-})$ and
$(M_{\ell^+ \ell^- \gamma}-M_{\ell^+ \ell^-})$,
to extract efficiency-corrected yields of feed-down $\po$.  
We thus obtain momentum distributions of $\po$ and $\pt$ 
which have been corrected for detection efficiency, 
$\po$ feed-down, and continuum background.
The yields are then normalized by 
$n_B \times {\cal B}(\psi \to \ell^+ \ell^- (\pi^+ \pi^-))$, where 
$n_B$ is the number of $B$ and $\overline{B}$ mesons in the data;
the uncertainties in these quantities are reflected in our results as an
overall scale factor error.
Inclusive branching fractions are obtained 
by summing the normalized momentum distributions over all bins.
Finally, the Monte Carlo simulation is used to obtain a matrix which 
correlates the momentum of the $\psi$ as generated 
to the momentum as measured;  
by inverting this matrix and applying it to the observed momentum 
distribution, we are able to deconvolve the effects of 
detector measurement smearing from the distribution.

We investigate the possible sources of systematic error;  
for each source, we make an appropriate modification to the 
measurement procedure and observe the deviation of the resulting yield
relative to the nominal procedure.
The deviations are then combined to obtain final systematic errors.
The sources of error are grouped as follows:
(1) Monte Carlo simulation of track and shower finding, 
electron and muon identification, $\psi$ polarization,
global event and kinematic cuts, 
(2) invariant mass fit procedure;
(3) branching fractions of unmeasured modes; and
(4) overall scale factor.

\begin{figure}
\includegraphics[height=5.208in,width=5.0625in]{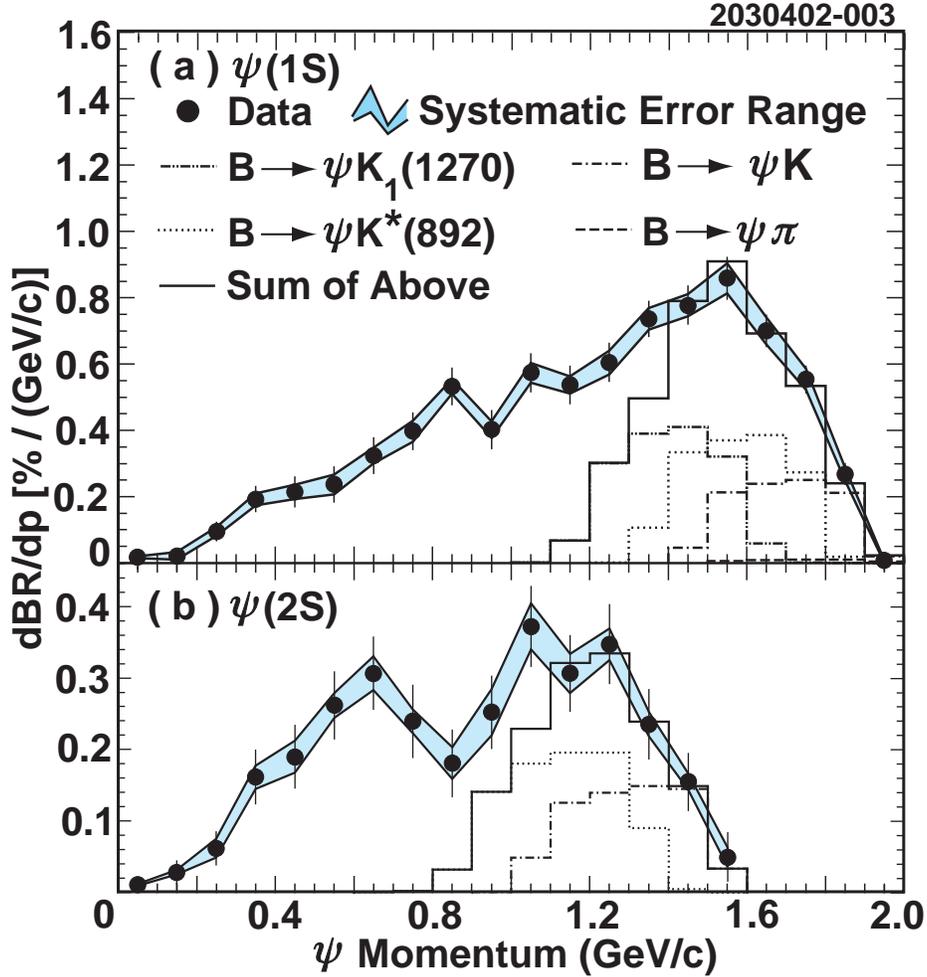}
\caption{\label{fig:p} 
Momentum distributions of (a) $\po$ and 
(b) $\pt$ produced directly from $B$ decays.  
There is an additional overall scale uncertainty of
$1.2\%$ for $\po$ and $5.1\%$ for $\pt$ which is not depicted in the plots.
The histograms show the contributions of two-body $B \to \psi X$ decays, 
where the lineshapes are obtained from Monte Carlo simulation 
and the normalizations are from previous 
determinations of exclusive branching fractions
\cite{PDG2001,Bellepsik1,CLEOpsi2k}.
}
\end{figure}

The results for the inclusive branching fractions are given in 
Table \ref{tab:BF}
and the momentum distributions are shown in Figure \ref{fig:p}. 
Our branching fraction results are consistent with 
previous published measurements \cite{CLEOBigB} 
as well as preliminary measurements \cite{BaBarBF}, 
and are limited by systematic errors.  The combined error for
${\cal B}(B \to \po(\mbox{direct}) X)$ is smaller than the error of the PDG
2001 average \cite{PDG2001} by a factor of two.
However, the theoretical uncertainties in the NRQCD calculations
of the branching fractions are such that the improved accuracy of this 
measurement is unlikely to further constrain the NRQCD LDME's.
The momentum distributions reported here are the first to subtract the 
{\it measured} distributions of feed-down and continuum $\psi$, 
correct for detector measurement smearing, and analyze systematic errors 
for each bin individually.  
Figure 1(b) is also the first to show the momentum distribution of
multibody ($\ge 3$-body) decays in $B \to \pt X$ production;  
these decays account for much of the total $B \to \pt X$ production, 
as is also the case with $B \to \po X$ \cite{Beneke97}.
It should be possible to update previous phenomenological studies 
of the $\psi$ momentum distribution with these improved
measurements.

\begin{table}[ht]
\begin{center}
\begin{tabular}{l c c}
\hline \hline
Decay & Branching Fraction (\%) \\ \hline
$B \to \po X$                & $1.121 \pm 0.013 \pm 0.040 \pm 0.013$  \\
$B \to \po(\mbox{direct}) X$ & $0.813 \pm 0.017 \pm 0.036 \pm 0.010$  \\
$B \to \co X \to \po X$      & $0.119 \pm 0.008 \pm 0.009 \pm 0.001$  \\
$B \to \co X$                & $0.435 \pm 0.029 \pm 0.031 \pm 0.026$  \\
$B \to \pt X \to \po X$      & $0.189 \pm 0.010 \pm 0.018 \pm 0.002$  \\ 
$B \to \pt X$                & $0.316 \pm 0.014 \pm 0.023 \pm 0.016$  \\  
\hline \hline
\end{tabular}
\end{center}
\caption[h] {
Inclusive branching fraction results.  
The errors shown are (in order) statistical, systematic, 
and due to an overall scale factor uncertainty.
\label{tab:BF}
}
\end{table}
\nopagebreak

The polarization parameter $\alpha$ is equal to ($+1$, 0, $-1$) 
for a population of (transversely, randomly, longitudinally) polarized $\psi$.
For $\psi \to \ell^+ \ell^-$ decays, it is determined experimentally by
measuring the decay angle $\theta$, which is defined as the angle between
the $\ell^+$ direction in the $\psi$ rest frame and the $\psi$ direction
in the $B$ rest frame.  
The $\cost$ distribution for a population of $\psi$ is proportional
to $(1 + \alpha \cos^2\theta)$.
The angular distribution is obtained in a similar manner as the 
momentum distributions:  the dataset is partitioned into 5 equal bins in 
$\cost$ between $-1$ and $1$;
for each partition, we fit the invariant mass distribution to
find the signal yield.
In addition to measuring the polarization of direct $\po$ and $\pt$ 
for all momenta, 
we also extract $\alpha_{\po}$ in 3 coarse momentum bins.

\begin{figure}
\includegraphics[height=2.7855in,width=5.0625in]{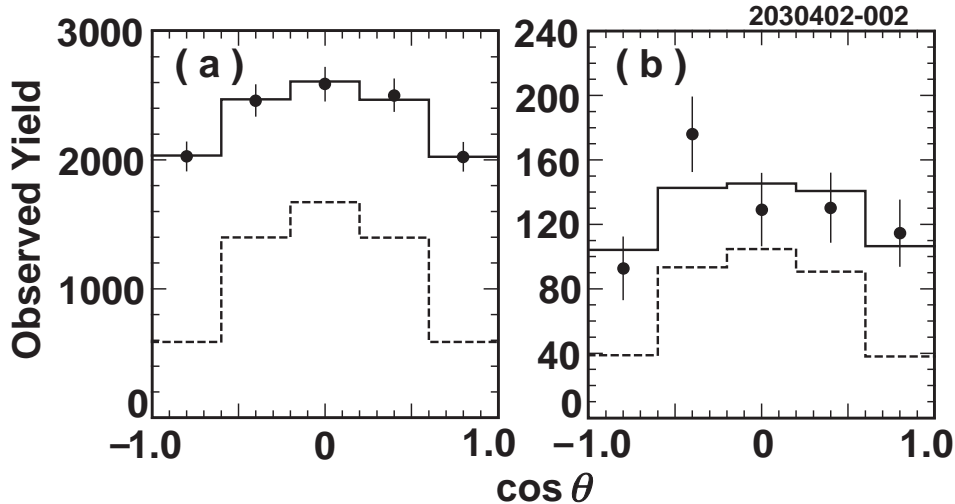}
\caption{\label{fig:angular} 
Decay angle distributions of (a) $\po$ and (b) $\pt$ 
from $B \to \psi\mbox{(direct)}X$, summed over all $p_\psi$.
The points represent the data, showing $1\sigma$ statistical errors.
In both figures, the fit result (solid histogram) 
is the sum of a longitudinal component (dashed histogram) 
and a transverse component.
}
\end{figure}

At CLEO, $B$ mesons are produced 
with a small boost in the $\Upsilon(4S)$ (lab) frame, 
the direction of which is unknown.
The boost of the $B$ results in a smeared measurement of $\cost$;  
directly fitting for $\alpha$ using the measured $\cost$ distribution 
would yield a biased result.
However, this kinematic smearing is accurately modeled
by the Monte Carlo simulation.  Our procedure is to generate Monte Carlo 
events in two sets; one with all $\psi$ generated longitudinally, 
the other with all transverse.  
The measured $\cost$ distribution from the
data is then fit to a sum of the reconstructed $\cost$ distributions 
from the polarized Monte Carlo sets.  This procedure correctly accounts
for both the boost smearing and detection efficiency.
Since the efficiency also depends on $p_\psi$, we must ensure that the
Monte Carlo distributions of generated $p_\psi$ match those of Figure
\ref{fig:p};  this is accomplished through a rejection technique.
Because the observed $\cost$ distributions are not directly corrected for 
detection efficiency, the observed feed-down distributions are 
corrected only for the efficiency of detecting the additional particles 
needed to reconstruct the $\pt$ or $\co$.
The final feed-down and continuum-subtracted angular distributions
are shown in Figure \ref{fig:angular}.
The systematic error study included the previously mentioned sources of bias;
additionally, we have investigated the possible systematic error arising for 
the 
methods for feed-down subtraction and fitting for $\alpha$.

The final polarization results are listed in Table \ref{tab:alpha};
these are the first results for the polarization of $\po$ and $\pt$ 
from $B \to \psi\mbox{(direct)} X$.  
For comparison, we measure $\alpha = -0.35 \pm 0.03$ (statistical error only)
for $\po$ from $B \to \po\mbox{(all)} X$.
Our result for $\alpha_{\po}$ is about $4$ standard deviations
from zero;  this measurement therefore strongly disfavors 
the color evaporation model of charmonium production \cite{Fritzsch77}, 
which predicts zero net polarization, independent of the production mechanism. 
When next-to-leading-order calculations become available, these results
also have the potential to significantly constrain the 
long-distance matrix elements of NRQCD.

We gratefully acknowledge the effort of the CESR staff in providing us with
excellent luminosity and running conditions.
This work was supported by 
the National Science Foundation,
the U.S. Department of Energy,
the Research Corporation,
and the Texas Advanced Research Program.

\renewcommand{\arraystretch}{1.5}
\begin{table}[ht]
\begin{tabular}[c]{@{\hspace{0.65cm}}c@{\hspace{0.7cm}}c@{\hspace{0.7cm}}c@{}c@{}c@{\hspace{0.65cm}}}
\hline \hline
$\psi$ Meson & $p_{\psi}$ (GeV/$c$) & \multicolumn{3}{c@{\hspace{0.65cm}}}{$\alpha$} \\ \hline
  $\po$ & $0.0 - 2.0$ & $-0.30$ & $^{+0.07}_{-0.06}$ & $\pm 0.04$ \\ 
  $\pt$ & $0.0 - 1.6$ & $-0.45$ & $^{+0.22}_{-0.19}$ & $\pm 0.04$ \\ \hline
  $\po$ & $0.0 - 0.8$ & $+0.32$ & $^{+0.33}_{-0.27}$ & $\pm 0.15$ \\ 
  $\po$ & $0.8 - 1.4$ & $-0.37$ & $^{+0.09}_{-0.09}$ & $\pm 0.04$ \\ 
  $\po$ & $1.4 - 2.0$ & $-0.52$ & $^{+0.08}_{-0.07}$ & $\pm 0.03$ \\
\hline \hline
\end{tabular}
\caption[h] {
Polarization of $\po$ and $\pt$ from $B \to \psi\mbox{(direct)}X$
over the full momentum range (top two values) 
and for $\po$ in three momentum ranges.  
The errors are statistical and systematic.
\label{tab:alpha}
}
\end{table}
\renewcommand{\arraystretch}{1.0}

\pagebreak

\bibliography{text}

\end{document}